\documentclass[11pt]{article}
\RequirePackage[OT1]{fontenc}
\usepackage{amsthm,amsmath,amssymb,amsfonts,bm,enumerate,xcolor,graphicx,natbib,color,url}
\RequirePackage[colorlinks,citecolor=blue,urlcolor=blue]{hyperref}
\usepackage{hyperref}
\usepackage{subcaption} 
\usepackage{ulem}
\usepackage{pifont}
\usepackage{textcmds}
\usepackage{booktabs}
\usepackage{multirow}
\usepackage{booktabs}
\usepackage{adjustbox}
\usepackage[usestackEOL]{stackengine}
\usepackage[toc,page]{appendix} 
\usepackage{algorithm,algorithmic}
\usepackage{lipsum}
\usepackage{setspace}

\usepackage{setspace}
\allowdisplaybreaks

\definecolor{lb}{RGB}{44,139,183}
\newcommand\cl[2]{\textcolor{#1}{#2}}
\newcommand\bcl[2]{\textcolor{#1}{{\fontseries{b}\selectfont #2}}}
\usepackage{ulem} 

\def\frac#1#2{{\textstyle{#1\over#2}}}

\DeclareSymbolFont{AMSb}{U}{msb}{m}{n}
\DeclareMathSymbol{\Natural}{\mathbin}{AMSb}{"4E}
\DeclareMathSymbol{\Integer}{\mathbin}{AMSb}{"5A}
\DeclareMathSymbol{\Real}{\mathbin}{AMSb}{"52}
\DeclareMathSymbol{\Rational}{\mathbin}{AMSb}{"51}
\DeclareMathSymbol{\Imaginary}{\mathbin}{AMSb}{"49}
\DeclareMathSymbol{\Complex}{\mathbin}{AMSb}{"43} 
\DeclareMathSymbol{\Disk}{\mathbin}{AMSb}{"44} 
\def\bi{\begin{itemize}}
\def\ei{\end{itemize}}
\def\bd{\begin{description}}
\def\ed{\end{description}}
\def\ben{\begin{enumerate}}
\def\een{\end{enumerate}}
\def\bv{\begin{verbatim}}
\def\ev{\end{verbatim}}

\def\bar#1{{\overline{#1}}}
\def\hat#1{{\widehat{#1}}}

\def\2to{{\ {\buildrel 2\over \longrightarrow}\ }}

\def\I1ton{{$I_1,\ldots,I_n$}}
\def\X1ton{{$X_1,\ldots,X_n$}}
\def\Y1ton{{$Y_1,\ldots,Y_n$}}
\def\Z1ton{{$Z_1,\ldots,Z_n$}}
\def\R1ton{{$R_1,\ldots,R_n$}}
\def\e1ton{{$e_1,\ldots,e_n$}}
\def\t1ton{{$t_1,\ldots,t_n$}}
\def\x1ton{{$x_1,\ldots,x_n$}}
\def\y1ton{{$y_1,\ldots,y_n$}}
\def\z1ton{{$z_1,\ldots,z_n$}}

\usepackage[english]{babel}
\usepackage{amsfonts}
\usepackage{amssymb}
\usepackage{graphicx}
\usepackage{enumerate}

\newcommand{\blind}{1}
\renewcommand{\arraystretch}{2}

\begin{document}
\thispagestyle{empty}
\baselineskip=28pt
\vskip 5mm

\begin{center} {\Large{\bf Joint modeling of landslide counts and sizes using spatial marked  point processes with sub-asymptotic mark distributions}}
\end{center}

\baselineskip=12pt

\vskip 5mm

\if1\blind
{
\begin{center}
\large
Rishikesh Yadav$^1$, Rapha\"el Huser$^1$, Thomas Opitz$^2$,  Luigi Lombardo$^3$\\ 
\end{center}
\footnotetext[1]{
\baselineskip=10pt Statistics Program, Computer, Electrical and Mathematical Sciences and Engineering (CEMSE) Division, King Abdullah University of Science and Technology (KAUST), Thuwal 23955-6900, Saudi Arabia. E-mails: rishikesh.yadav@kaust.edu.sa; raphael.huser@kaust.edu.sa}
\footnotetext[2]{
\baselineskip=10pt  INRAE, UR546 Biostatistics and Spatial Processes, 228, Route de l'A\'erodrome, CS 40509, 84914 Avignon, France. E-mail: thomas.opitz@inrae.fr}
\footnotetext[3]{Faculty of Geo-Information Science and Earth Observation (ITC), University of Twente, P.O. Box 217, Enschede, AE 7500, the Netherlands. E-mail: l.lombardo@utwente.nl }
} \fi

\baselineskip=26pt
\vskip 2mm
\centerline{\today}
\vskip 4mm



{\large{\bf Abstract}} To accurately quantify landslide hazard in a region of Turkey, we develop new marked point process models within a Bayesian hierarchical framework for the joint prediction of landslide counts and sizes. To accommodate for the dominant role of the few largest landslides in aggregated sizes, we leverage mark distributions with strong justification from extreme-value theory, thus bridging the two broad areas of statistics of extremes and marked point patterns. At the data level, we assume a Poisson distribution for landslide counts, while we compare different ``sub-asymptotic" distributions for landslide sizes to flexibly model their upper and lower tails. At the latent level, Poisson intensities and the median of the size distribution vary spatially in terms of fixed and random effects, with shared spatial components capturing cross-correlation between landslide counts and sizes. We robustly model spatial dependence using intrinsic conditional autoregressive priors. Our novel models are fitted efficiently using a customized adaptive Markov chain Monte Carlo algorithm. We show that, for our dataset, sub-asymptotic mark distributions provide improved predictions of large landslide sizes compared to more traditional choices. To showcase the benefits of joint occurrence-size models and illustrate their usefulness for risk assessment, we map landslide hazard along major roads.

\baselineskip=26pt


{\bf Keywords:} Bayesian hierarchical modeling;  extreme event; landslides hazard; marked point process; Markov chain Monte Carlo; sub-asymptotic modeling.


\baselineskip=26pt
\section{Introduction} 
\label{sec:introduction}

Landslides are a severe natural hazard worldwide and are common in mountains and hills where they can pose a severe threat to human lives, disrupt services such as water supply and destroy public and private properties, generating annual damages amounting to billions of dollars \citep{kennedy2015systematic, daniell2017losses, broeckx2020landslide}. It is essential to understand the physical mechanisms triggering devastating landslides, and to assess future landslide risk in terms of various geophysical, geomorphologic, thematic, or climatic factors that can be accurately measured. Susceptibility maps are the most common ways to predict the presence and absence of landslide occurrences over a fixed geographical region. These maps are generally spatial processes and often result from spatial predictive models \citep{brenning2005spatial, chen2017landslide}. Most of the current research focuses on mapping landslides by exploiting geographical covariates to predict presence-absence information \citep{ayalew2005application, goetz2015evaluating, camilo2017handling}; see also \cite{reichenbach2018review} for a recent review. However, no stochastic spatial dependence is introduced in these models, and the spatial information is introduced only through available covariates that vary over space. \cite{lombardo2018point} developed, for the first time, a Bayesian hierarchical model with desired spatial dependence structure based on the ``intensity'' concept for spatial landslides prediction; some other more recent examples include \citet{lombardo2019geostatistical,lombardo2019numerical,lombardo2020space}. Precisely, they proposed a Bayesian hierarchical model, where landslide counts are viewed as a spatial or spatio-temporal point pattern of log-Gaussian Cox type, and they used the integrated nested Laplace approximation \cite[INLA,][]{rue2009approximate} for the inference of such models. 
Recently, \cite{opitz2020high} used one of their models as a baseline  and explored more complex constructions involving small-scale random variations and space-varying regression to impose certain natural physical constraints and further enhance landslide predictions. In their framework, high-resolution mapping of landslides was made possible by introducing random effects at a physically-defined lower resolution, using so-called slope units \citep[SU,][]{amato2019accounting}, while covariate information was still used at the higher spatial resolution (pixels). However, while this flexible class of models is helpful to predict landslide occurrences and counts per mapping unit, this methodology still completely ignores the actual destructiveness of the landslides themselves, which is perhaps even more important for landslide risk assessment and country planning.

Landslide hazard, by definition, is directly linked to  actual landslide size \citep{guzzetti1999, tanyas2019}, be it expressed as the volume of the displaced mass, or the area of the landslide scar,
or indirectly to the landslide ``diameter" and other length-to-width ratio properties \citep{taylor2018}. Therefore, it is crucial to study the landslide size distribution jointly with landslide occurrences. Here, we shall define the landslide size by (a function of) the landslide planimetric area, which is relatively easy to measure from remotely sensed images and serves as a good proxy for the landslide volume, thanks to a well-known area--volume conversion formula. Precisely, we shall model the square root of the landslide area, which is lighter-tailed than the area itself, and can be measured and intuitively interpreted on the same scale as the landslide diameter. In the literature, other attempts have already been made to model and predict landslide sizes or ``magnitudes'' in addition to the modeling of landslide counts \citep[see, e.g.,][]{guo2017size, roback2018size, valagussa2019seismic, vanani2021statistical, lombardo2021landslide}. For example, \cite{guo2017size} used power-law relationships of the size distribution to study earthquake-induced landslides in both the Himalayan and Lesser Himalayan regions. In some related work, \cite{vanani2021statistical} studied the correlation between the landslide counts and sizes using a bivariate model and a linear automated modeling procedure using the SPSS software. More recently, \citet{lombardo2021landslide} proposed a statistically-based model to estimate the area of landslides aggregated over slope units, though they completely ignored their occurrence probability. Specifically, they proposed a Bayesian version of generalized additive models where both the maximum landslide size per slope unit and the sum of all landslide sizes per slope unit were modeled using a log-normal model. In this paper, we propose instead using marked point process models for the joint modeling of landslide counts and sizes at high resolution, with various mark distributions strongly justified by extreme-value theory. At the data level, landslide counts are assumed to independently follow a Poisson distribution with spatially varying intensity function, and landslide sizes are assumed to independently follow a sub-asymptotic extreme-value distribution with spatially varying median process. The sub-asymptotic distributions used in this work have been proposed in recent literature \citep{Papastathopoulos.Tawn:2013,Naveau.etal:2016, yadav2019spatial, yadav2021flexible} and are theoretically justified by extreme-value theory; see also \citet{Stein:2021a,Stein:2021b} who recently developed a constructive framework and new models for sub-asymptotic distributions with flexible behavior in both tails. As these models are specifically designed to be flexible both in their lower and upper tails, they provide a natural framework for accurately modeling landslide size data all the way from low to high quantiles. This is important in our landslide hazard assessment context, because the largest landslides are usually the most destructive ones and need to be modeled accurately using tail-focused models, while smaller landslides can still in some cases lead to important damages and should not be ignored. All of the sub-asymptotic distributions that we use in this work directly extend the generalized Pareto (GP) distribution \citep{Davison.Smith:1990}, commonly used as an asymptotic model for high threshold exceedances. Since the GP distribution can be obtained as a special case of these sub-asymptotic distributions, model parameters thus have an intuitive interpretation, and our framework allows us to assess ``how far'' the estimated landslide size distribution is from its upper tail limit behavior.

In other applied contexts, related approaches have been proposed for the joint statistical modeling of occurrence and size data, sometimes treating sizes as numerical ``marks'' of the occurrence positions, e.g., see \citet{tonini2017evolution}, \citet{pimont2021prediction}, \citet{koh2021spatiotemporal}, and \citet{daniela2021jointmodel} for wildfire count and burnt area modeling. 
In particular, assuming a marked spatio-temporal log-Gaussian Cox process for daily wildfire data in France, \cite{pimont2021prediction} and \cite{koh2021spatiotemporal} developed a joint Bayesian hierarchical model within the INLA framework exploiting extreme-value theory, but sub-asymptotic mark distributions were not considered. Instead, \cite{koh2021spatiotemporal} modeled the size distribution using so-called ``split-models'': extreme threshold exceedances were assumed to follow the GP distribution, and moderate values followed a rescaled Beta distribution. Similar split-models were introduced by \cite{patel2021spatio} to distinguish bulk and tail properties of marks in a self-exciting marked spatio-temporal point process for Afghanistan terror attacks. 
By contrast, our new models 
avoid split-modeling and instead rely on sub-asymptotic extreme-value distributions, which contain the asymptotic GP distribution as a special case, while complying with extreme-value theory in both tails. 
Our approach has the great benefit of providing a parsimonious and unified model for the full mark distribution with continuous density, thus bypassing the tricky threshold choice and the unrealistic model discontinuity at the threshold. We note that a parsimonious model for the mark distribution is crucial to accurately model relatively small spatial marked point process datasets. Therefore, it is important to use relatively simple distributions that are well-supported theoretically, have interpretable model parameters, and remain flexible in the way extremes are captured. These arguments provide strong support for sub-asymptotic distributions, and in this work, we test different families of sub-asymptotic distributions and compare them with more classical choices (e.g., log-normal, Gamma, etc.). 

The Bayesian hierarchical framework allows for explanatory and predictive modeling, where  Gaussian prior distribution are often assumed at the latent level to capture non-linear effects of covariates and spatial coordinates along with spatial dependence \citep{Banerjee.etal:2014, Cressie:1993}. Our marked point processes possess latent Gaussian processes that are combinations of fixed effects, spatial random effects assumed to follow intrinsic conditional autoregressive (ICAR) priors \citep{besag1975statistical}, and independent Gaussian random effects. In contrast to INLA-based approaches inheriting the implementation restrictions of INLA, our Bayesian inference procedure exploits a general and efficient Markov-chain Monte-Carlo (MCMC) sampling scheme, with customized updates for our proposed random effect modeling framework.
It combines standard Metropolis--Hastings updates \citep{metropolis1953equation}, Gibbs sampling \citep{casella1992explaining}, and the adaptive Metropolis adjusted Langevin algorithm (MALA) developed in \cite{yadav2019spatial} to enhance its efficiency with numerous latent Gaussian variables and hyperparameters. Although our MCMC-based inference is computationally more demanding than other approximate Bayesian inference techniques such as INLA, it has two major advantages: (i) MCMC samplers are known to provide ``pseudo-exact" inferences, provided convergence of Markov chains is fast enough and the algorithm is run for a sufficient number of iterations; and (ii) new types of mark distributions or model structures can be readily incorporated unlike for INLA, which
requires certain distributional properties (e.g., log-concave densities with respect to the latent linear predictor) to work correctly. 

The manuscript is organized as follows. In \S\ref{sec:DataDescr}, we describe available data and covariates. 
In \S\ref{sec:ModFraeme}, we specify our Bayesian hierarchical models.
\S\ref{sec:BayesInfSim} explains the Bayesian inference scheme and contains a simulation study to validate accurate MCMC performance. In \S\ref{sec:Chapter5Application} we illustrate our new methodology on the landslide dataset from Turkey. Specifically, we fit different models, especially for different mark distributions, and compare different models through cross-validation criteria. Concluding remarks and  further research directions are given in \S\ref{sec:Conclsuion}.

\section{Landslide data and predictor variables}
\label{sec:DataDescr}
\subsection{Study area and data description}
\begin{figure}[t!]
\includegraphics[width=\linewidth]{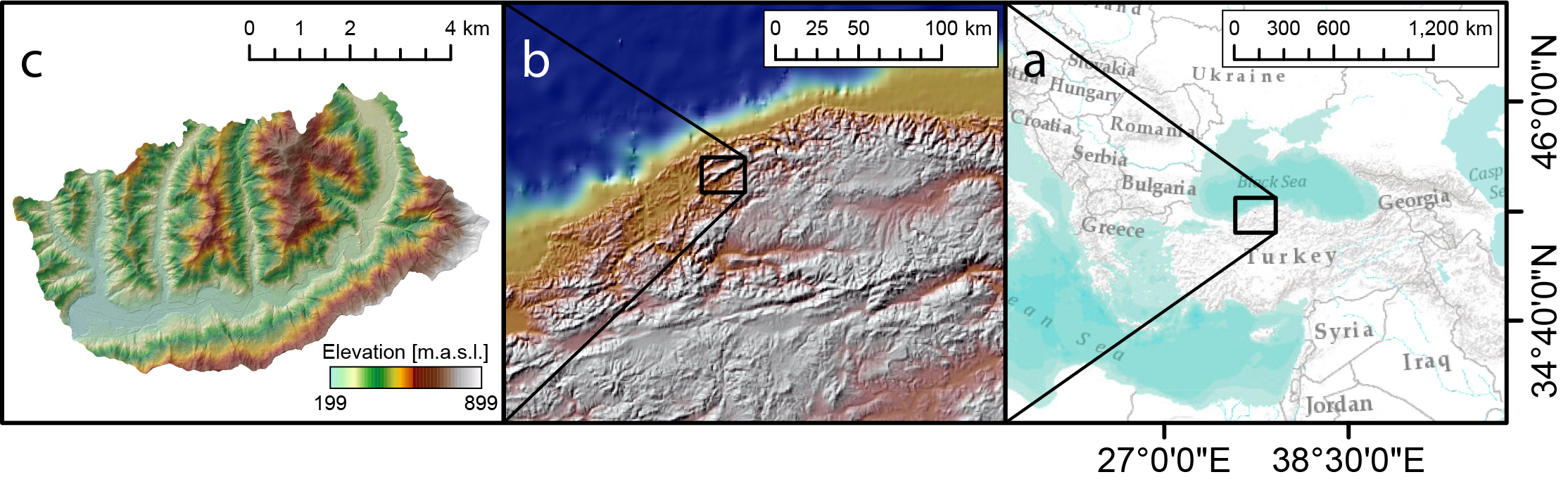}
\caption{a) Large geographic context; b) Zoom into the area surrounding the study zone, showing its proximity to the Black Sea in the north, responsible for summer cloudbursts, and to large fault lineaments in the south, responsible for tectonic deformation; c) Study area shown in terms of elevation superimposed on the shaded relief.}
\label{fig:Fig1}
\end{figure}

The study area covers $33.2$ km$^2$ in the municipality of Ulus (Bartin) in the Western Black Sea, Turkey, drained by the Ulus river. It  lies between the following geographical coordinates: $41^{\circ}34'39''$N--$41^{\circ}34'35''$N;  $32^{\circ}37'20''$E--$32^{\circ}43'31''$E.  Figure~\ref{fig:Fig1} shows a topographic map of the study region.
The landslide inventory is based on airborne Light Detection and Ranging (LiDAR) technology. Elevation within the study region varies from 195m to 900m with a mean  of 375m, computed from a 1m $\times $ 1m digital elevation model (DEM). The Ulus region is mostly hilly and has a maritime climate with mean annual precipitation of 1020mm. Landslides mostly occur as a result of intense and prolonged rainfall events \citep{can2005susceptibility}. For more details,
see \cite{gorum2019landslide}. 
\begin{figure}[t!]
\includegraphics[width=\linewidth]{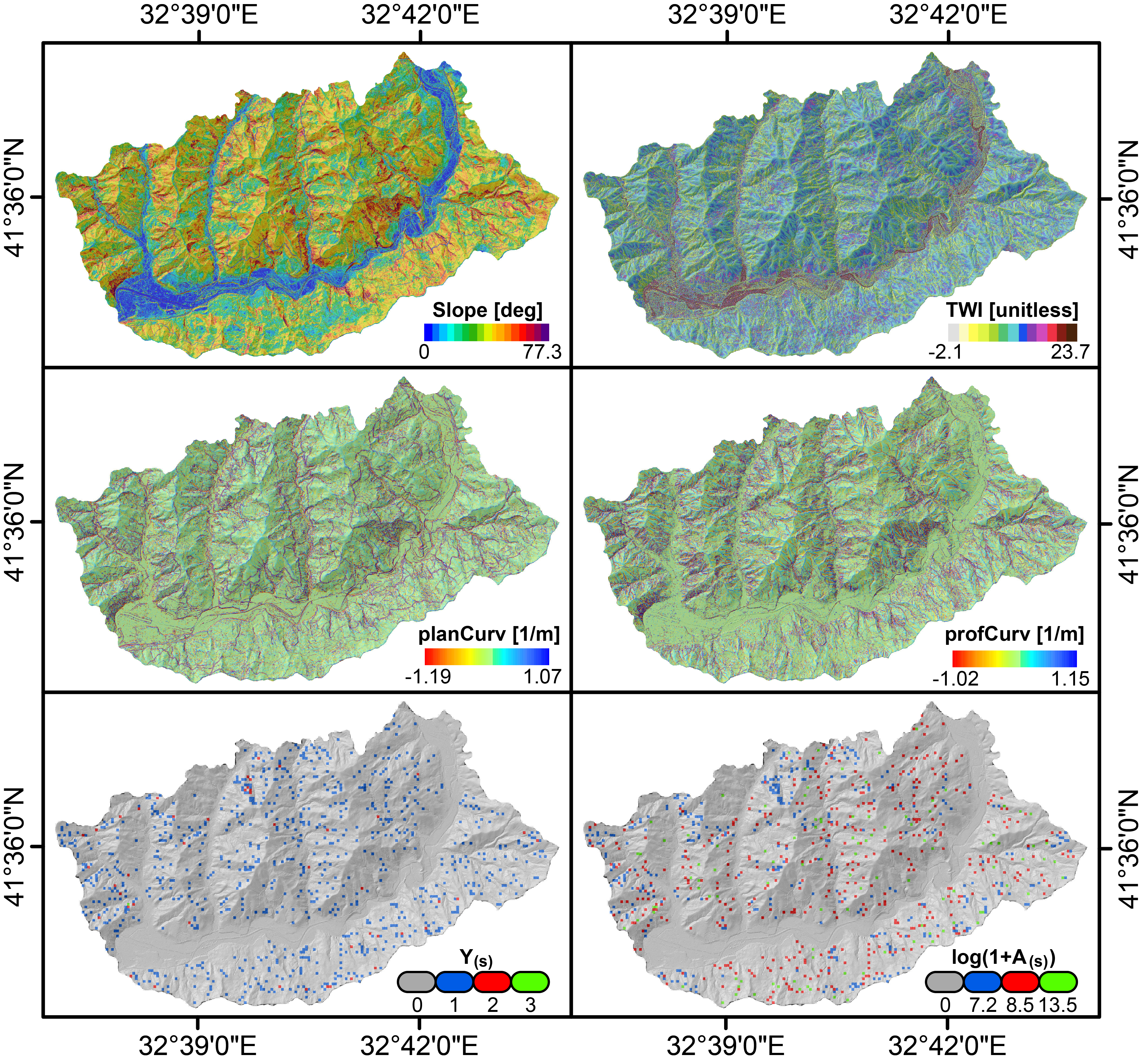}
\caption{Example of four covariates (details in \S\ref{subsec:CovDescr}) at the level of the LiDAR survey: Slope Steepness (top left), Topographic Wetness Index (top right), Planar and Profile Curvatures (middle left and right). Bottom row: number of landslides (left) in a $50 \times 50$~m lattice; and landslide areas (right; on log-scale) for each landslide occurrence.}
\label{fig:DataDes}
\end{figure}
Figure~\ref{fig:DataDes} shows landslide locations (bottom left panel) and their landslide areas ($m^2$) in log-scale (bottom right panel) along with important predictor variables, such as the slope,  detailed in \S\ref{subsec:CovDescr}.
Regions with lower slopes tend to have lower landslide counts and sizes. Landslide counts and sizes are available at high spatial resolution (pixels) with a pixel area of 50m $\times$ 50m. In total, 933 landslides are observed in the study region, where 836 pixels contain a single landslide, 47 pixels contain two landslides, and  one pixel contains three landslides.

\subsection{Covariate information}
\label{subsec:CovDescr}
\begin{figure}[t!]
\includegraphics[width=\linewidth]{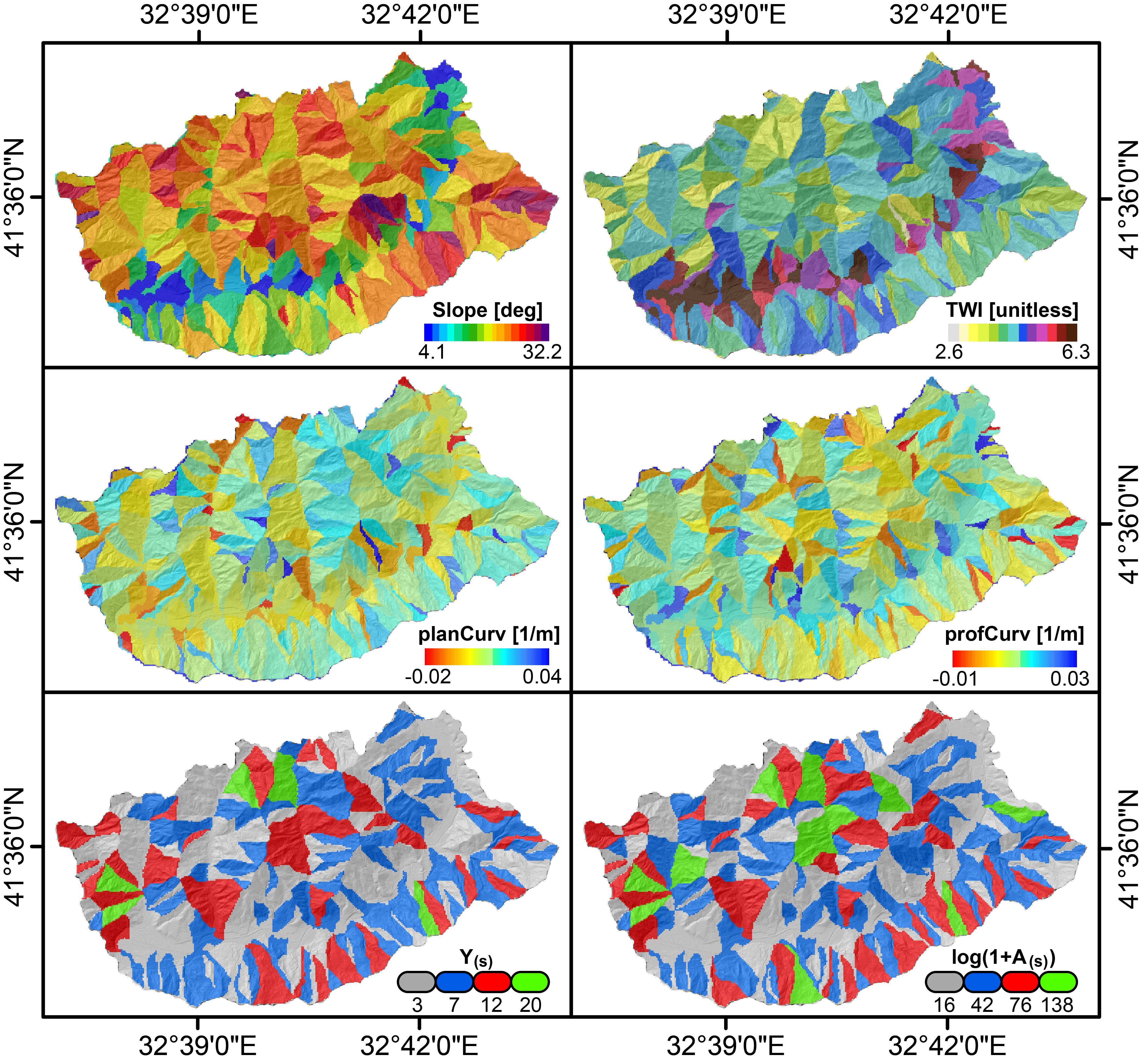}
\caption{Example of four covariates aggregated to the mean value per slope unit: Slope Steepness (top left), Topographic Wetness Index (top right), Planar and Profile Curvatures (middle left and right). Bottom row: number of landslides (left) and associated landslide areas (right) displayed on log-scale and aggregated at the slope unit level.}
\label{fig:CountsSize_SU}
\end{figure}

Nine morphometric covariates are available for every pixel, each observed at the  pixel level of the LiDAR survey and 
consisting of a series of elevation derivatives computed from a 1m $\times$ 1m DEM, namely slope \citep{zevenbergen1987quantitative}, vector ruggedness measure \citep[VRM;][]{sappington2007quantifying}, planar and profile Curvatures \citep[planCurv and profCurv, respectively;][]{heerdegen1982quantifying}, Topographic Positioning Index \citep[TPI;][]{de2013}, Topographic Wetness Index \cite[TWI;][]{kirkby1979physically}, local relief \citep[LR;][]{stepinski2011}, slope height \citep[s-height;][]{evans2019} and valley depth \citep[v-depth;][]{loczy2012}. Most of these covariates have already been shown to be associated with the landsliding process; see for example \cite{lombardo2018point}, and references therein. Four of these covariates (slope, TWI, planCurv and profCurv) are graphically illustrated in Figures~\ref{fig:DataDes} and \ref{fig:CountsSize_SU}.

\subsection{Mapping units}
We consider two types of mapping units covering the entire study region: (i) a high-resolution regular grid (pixels) comprising 13375 squared pixels (4 pixels are deleted in the final analysis due to some outliers in covariate values), each having an approximate size of 2500m$^2$, and (ii) a slope unit \cite[SU,][]{amato2019accounting} partition, which is at a lower resolution than pixels but is defined in a meaningful physically-based way that is relevant to the landsliding process; see, e.g., \cite{lombardo2018point}. We have  355 slope units in our study region. 

In our marked point process models, we express landslide counts, sizes, and covariates on the fine spatial grid resolution (pixel level), see Figure~\ref{fig:DataDes}, whereas the latent spatial effects are defined at the coarser slope unit resolution to reduce the computational burden;  \S\ref{subsec:MarkedPoint-pixel} provides more details on this modeling approach. Defining random effects at high resolution (pixel level) is computationally demanding, and the strong dependence between adjacent pixels might increase numerical instabilities. Instead, slope units are attractive for their physical interpretation because they are geomorphologically independent spatial entities that provide a homogeneous response of a given slope when a landslide occurs, and they also lead to considerably faster and more stable inference because of their coarser resolution. For these reasons, we thus adopt a hybrid modeling approach, with landslide data and covariates at high pixel resolution, and random effects at lower slope unit resolution. An alternative modeling strategy is to aggregate landslide counts, sizes, and covariates at the slope unit level, and to build a joint areal model; see \S\ref{subsec:ArealModel} for details. Figure~\ref{fig:CountsSize_SU} shows an example of covariates as well as the landslide counts and areas calculated at the slope unit scale; all covariates  are obtained by averaging, and counts and sizes are obtained by adding up the corresponding pixel-based values. While we here mostly focus on marked point process models, Section~\ref{subsec:arealresults} summarizes results from joint areal models, and the Supplementary Material provides further details and a comparison with marked point process models.

\section{Modeling framework}
\label{sec:ModFraeme}
\subsection{Log-Gaussian Cox processes}
Log-Gaussian Cox processes \cite[LGCP,][]{moller1998log} are Poisson processes with random intensity function given by a log-Gaussian process. Thanks to their doubly stochastic structure, these processes are convenient for modeling the spatial clustering of points due to unobserved environmental predictors. In Bayesian hierarchical modeling,  latent processes are viewed as Gaussian process priors on the Poisson log-intensity function and may contain various fixed and random effects that are aggregated additively to preserve the Gaussian distribution of the linear predictor, and therefore the LGCP structure. To hierarchically define the LGCP, let $\Lambda (\bm s)$ denote the random intensity function of the point process, and let $Y(B)$ denote the random counting process indicating the number of points  in a Borel set $B\subset \mathcal{S}$ contained in the study area $\mathcal{S}\subset\mathbb{R}^2$. At the data level, we have
\begin{align}
\label{eq:LGCP}
Y(B) \mid \Lambda(\bm s)\sim {\rm{Poisson}} \left\{ \int_B\Lambda(\bm s)\,\mathrm{d}\bm s\right\}.
\end{align}
We write   $\eta(\bm s)$ for the log-intensity function  into which we additively include jointly Gaussian spatial structures via $K$ random effects $x_{k}$, $k=1,\ldots, K$, $P$ fixed effect covariates $z_p$ with corresponding coefficients $\beta_p$, $p=1,\ldots,P$, and  an  intercept $\beta_0$, i.e.,
\begin{align*}
 \eta(\bm s)=\log\Lambda(s)=\beta_0+\sum_{p=1}^{P} \beta_p z_{p}(\bm s)+\sum_{k=1}^{K}x_k(\bm s), \qquad \bm s\in \mathcal{S}.
\end{align*}

Poisson processes and LGCPs are widely used in the modeling of spatial point patterns and have risen as a counterpart of Gaussian processes (used for continuous variables) for modeling discrete spatial phenomena. Recently, LGCP models have drawn  attention in modeling landslide occurrences; see, e.g., \citet{lombardo2018point, lombardo2020space, lombardo2021landslide} and \citet{opitz2020high}. In this work, we take LGCP models as a basis and in \S\S\ref{subsec:MarkedPoint-pixel}--\ref{subsec:ArealModel}, we build joint models for landslide occurrences and sizes (here defined as the square root of landslide areas), where sizes are viewed as numerical marks, each associated with exactly one point. Stochastic dependence could arise between the mark distribution of sizes and the random intensity function of occurrences, and our models will be designed to capture such dependence. In practice, the spatial domain $\mathcal{S}$ is often finely discretized for statistical inference on  LGCP models, for instance through a fine regular pixel grid, where the intensity function is assumed constant within each grid cell. Therefore, given the grid cells $B_i$, we rewrite the LGCP model as a sample of conditionally independent Poisson variables $N(B_i)$, $i=1,\ldots,n$, with random mean that follows a spatially correlated log-Gaussian process, and which provide grid-cell-based landslide counts according to \eqref{eq:LGCP}.  Small rectangular pixels may be replaced by more general choices of less regular and sometimes larger mapping units to discretize $\mathcal{S}$, for instance physically-based slope units, though at the cost of degrading the approximation of the continuous-space point process.

\subsection{Marked point process models}
\label{subsec:MarkedPoint-pixel}
We use generic notation $Y(\bm s)$ and $A(\bm s)$ for counting and size processes, respectively, observed at the spatial location $\bm s$ (or within the grid cell surrounding its representative point $\bm s$) in the spatial domain $\mathcal{\bm S}\subset \mathbb{R}^2$.  This means  $Y(\bm s_i)$ indicates the landslide occurrences over grid cell $i \in \{1,2,\ldots,n_1\}$ with centroid $\bm s_i$. We denote the surface area of grid cell $i$ by $e_i>0$. Given a positive landslide count $Y_i=Y(\bm s_i) >0$, we define $\bm A(\bm s_i)=\{A_{1}(\bm s_i),\ldots, A_{Y_{i}}(\bm s_i)\}\in (0, \infty)^{Y_i}$ as the continuous mark vector gathering the landslide sizes at the grid cell $i$.  Then, we define our general marked point process model as follows:
\begin{align}
\label{eq:GenModel}
Y(\bm s_i) \mid \eta(\bm s_i) &\stackrel{\rm {ind}}{\sim}  \text{Poisson}\left[e_i\exp\left\{\eta(\bm s_i)\right\}\right],\quad i=1,\ldots,n_1;\\
\eta(\bm s_i) &=\gamma_1 + \sum_{p=1}^{P_1} \beta_{1p} Z_{1p}(\bm s_i) \bm+\bm a_1(\bm s_i)^T\bm W_1 + \varepsilon_{\bm\eta}(\bm s_i); \nonumber \\
A_{j}(\bm s_l) \mid \mu_{j}(\bm s_l) &\stackrel{\rm {ind}}{\sim} f_A[\,\cdot\,;\,\exp\{\mu_{j}(\bm s_l)\},\bm\Theta_A],\quad l: Y(\bm s_l)>0; \quad j=1,\ldots,Y(\bm s_l);\nonumber \\
\mu_{j}(\bm s_l) &=\gamma_2 + \sum_{q=1}^{P_2} \beta_{2q} Z_{2q}(\bm s_l)+ \beta\bm a_2(\bm s_l)^T \bm W_1 + \bm a_2(\bm s_l)^T \bm W_2+ \varepsilon_{\bm\mu}(\bm s_l). \nonumber
\end{align}

Here, the processes $\bm W_1$, $\bm W_2$ are two spatially structured random vectors, while $\varepsilon_{\bm\eta}(\cdot)$, $\varepsilon_{\bm\mu}(\cdot)$ are independent (spatially unstructured) random effects, defined at the pixel level, included in the log-intensity process $\log\eta(\bm s)$, and log-median process $\log\mu(\bm s)$, respectively.  In particular, we assume that the spatial random effects $\bm W_1$ and $\bm W_2$ follow an intrinsic conditional autoregressive (ICAR) prior \citep{besag1975statistical, besag1995conditional}. More explicitly, $\bm W_1$ and $\bm W_2$ follow a singular multivariate Gaussian distribution with mean vector zero and precision matrices $\tau_{\bm w_1}Q$, and $\tau_{\bm w_2}Q$, respectively, where $\tau_{\bm w_1},\tau_{\bm w_2}>0$ are related to the marginal precision parameters. The matrix $Q$ is fixed for a given spatial structure and can be obtained as $Q=D-A$, where $D={\rm diag}(d_1,\ldots, d_N)$ is the diagonal matrix constructed from  the number of neighbors $d_i$ of region $i$, $A$ is the adjacency matrix between slope units,  and $N$ is the total number of spatial regions (here, slope units). The random effects  $\bm W_1$ and $\bm W_2$ are thus defined at the slope unit (SU) level (low resolution), while the fixed effects are used at the pixel level (high resolution). The vectors $\bm a_1(\bm s_i)$ and $\bm a_2(\bm s_l)$ project the information from the slope unit to the pixel level for count  and size processes, respectively, and are defined as: $\bm a_1(\bm s_i)=\{a_{11}(\bm s_i), \ldots, a_{1N}(\bm s_i)\}^T$, with $a_{1r}(\bm s_i)=1$ if the pixel $\bm s_i$ belongs to the $r^{\rm th}$ slope unit and $a_{1r}(\bm s_i)=0$ otherwise; similarly, $\bm a_2(\bm s_l)=\{a_{21}(\bm s_l), \ldots, a_{2N}(\bm s_l)\}^T$, with $a_{2r}(\bm s_l)=1$ if the pixel $\bm s_l$ belongs to the $r^{\rm th}$ slope unit, and $a_{2r}(\bm s_l)=0$ otherwise, where $l$ is such that $Y(\bm s_l)>0$. The observed covariate vector $\bm Z_{1p}=\{Z_{1p}(\bm s_1),\ldots, Z_{1p}(\bm s_{n{1}})\}^T$, $p=1,\ldots,P_1$, denotes  $P_1$ fixed effect covariates corresponding to the count process $Y(\bm s)$, and similarly $\bm Z_{2q}=\{Z_{2q}(\bm s_1),\ldots, Z_{2q}(\bm s_L)\}^T$, where $L=\sum_{i=1}^{n_1} Y(\bm s_i)$, $q=1,\ldots,P_2$, denotes $P_2$ fixed effect covariates used for the mark process $A(\bm s)$. The parameter vectors $\bm \beta_1=(\beta_{11},\ldots,\beta_{1P_1})^T$ and $\bm \beta_2=(\beta_{21},\ldots,\beta_{2P_{2}})^T$ are the corresponding covariate coefficients, and $\gamma_1$, $\gamma_2$ are the corresponding intercept parameters. The parameter $\beta$ scales the common component $\bm W_1$ and  controls how much information is shared from the count predictor towards the size predictor and determines the strength of interaction between the two processes. Precisely, it allows capturing positive ($\beta>0$) or negative ($\beta<0$) correlation. If $\beta=0$, then $Y(\bm s)$ is independent of $A(\bm s)$, and the model \eqref{eq:GenModel} for counts becomes the standard log-Gaussian Cox process model in \eqref{eq:LGCP}.

\subsection{Subasymptotic mark distributions}

We use the term \emph{subasymptotic distributions} for probability distributions applying to the full range of possible values of landslide sizes but possessing high flexibility, as required by extreme-value theory, for modeling and interpretation of both the lower and upper tail properties controlled by specific parameters.  These distributions 
provide different degrees of flexibility for separate control over bulk and tail features depending on their overall number of parameters.  
By carefully selecting the probability density $f_A\{\, \cdot\,;\mu(\bm s),\bm \Theta_A\}$ underlying the mark process $A(\bm s)$, we obtain models \eqref{eq:GenModel} with structurally different subasymptotic mark distributions; see Table~\ref{tab:DiffMarksDist} for certain choices of $f_A(\cdot)$ and their tail characteristics.  Various joint occurrence-size models are obtained by combining different mark distributions with different specifications of the structured spatial and independent random effects; see \S\ref{sec:Chapter5Application} for details. 
\begin{table}[t!]
\caption{Density functions of different mark distributions, reparametrized such that $\mu$ represents the median. Notation: incomplete gamma function $\Gamma(x)=\int_{0}^{x} t^{x-1}\exp(-t){\rm d}t$; incomplete beta function $B(x;\,a,b)=\int_{0}^{x} t^{a-1} (1-t)^{b-1} {\rm d}t$;  inverse of gamma cumulative distribution function (cdf) $F_{G}^{-1}(x;\, \, a,\, b)$ for $x\in [0,1]$ with shape $a$ and rate $b$; cdf $F_{Bu}^{-1}(x;\,c, \,\kappa, \lambda)$ of the Burr distribution with shape parameters $c,\kappa$ and scale parameter $\lambda$;   Asterisks ($^*$) indicate families of subasymptotic distributions that comply with extreme-value theory in both tails by allowing for any possible combination of positive upper-tail index $\xi_U$  and negative lower-tail index $\xi_L$; finally, $\kappa_U$ denotes the upper-tail Weibull index; $a_{+}=a$ if $a>0$ and $0$ otherwise; $\mathbb{I}$ is the indicator function. 
 }
\vspace{2mm}
\begin{adjustbox}{max width=\textwidth}
\begin{tabular}{c|c|c|c|c}
 marks & reparametrized density $f_A(x), x>0$ & $\bm \Theta_{A}$ & $\xi_U,\xi_L$ & $\kappa_U$
 \\
 \hline
gen-Gam & $f_{gG}(x)={{(c/\sigma^{\kappa})}\over{\Gamma(\kappa/c)}} {{x^{\kappa-1}}\over {\exp\left\{\left({{x}/{\sigma}}\right)^c\right\}}}$, $\sigma={{\mu}\over{\{F_{G}^{-1}(0.5;\, \kappa/c, \,1)\}^{1/c}}}$ &  $(\kappa>0, c>0)$ &$0, -1/\kappa$  & $c$\\
Gamma & $f_G(x)= f_{gG}(x)$ with $c=1$ & $\kappa>0$ &  $0, -1/\kappa$& $1$\\
Weibull & $f_{Wb}(x)=f_{gG}(x)$ with $c=\kappa$  & $\kappa>0$ & $0, -1/\kappa$ & $\kappa$ \\
l-Gamma & $f_{lG}(x)={{1}\over{\sigma^{\kappa}\Gamma(\kappa)}}  {{\left\{\log(x+1)\right\}^{\kappa-1}}\over{(x+1)^{1+1/\sigma}}}  $, $\sigma={{\log(1+\mu)}\over{ F_{G}^{-1}(0.5;\, \kappa, \,1)}}$   & $\kappa>0$ & $\sigma, -1/\kappa $  & $0$\\
l-Normal & $f_{lN}(x)={{\sqrt{\kappa}}\over{x\sqrt{2\pi}}}\exp\left[-{{\kappa}\over{2}} \{\log(x)-\sigma\}^2\right]$, $\sigma=\mu$ & $\kappa>0$ & $0,-\infty$  & $0$\\
Burr$^{*}$  & $f_{Bu}(x)={{(c\kappa/\sigma) (x/\sigma)^{c-1}}\over{\{1+(x/\sigma)^c\}^{\kappa+1}}}$, $\sigma={{\mu}\over{(2^{1/\kappa}-1)^{1/c}}}$ &  $(\kappa>0, c>0)$ & $1/c\kappa$, $-1/c$ & 0 \\
ext-GPD$^{*}$ & $f_{eGP}(x)=\begin{cases}
{{\kappa \left\{1-\left(1+\xi x/\sigma\right)^{-1/\xi}\right\}^{\kappa-1}}\over{\sigma\left(1+\xi x/\sigma\right)_{+}^{1+1/\xi}}}, & \xi > 0, \\ \sigma={{\mu \xi}\over{(1-0.5^{1/k})^{-\xi}-1}} \\ {{\kappa \left\{1-\exp\left(-x/\sigma\right)\right\}^{\kappa-1}}\over{\sigma \exp\left(x/\sigma\right)}}, & \xi\to 0, \\  \sigma={{\mu}\over{\log(1-0.5^{1/k})}}\\ \end{cases} $ & $ \kappa>0, \xi\geq 0$ &  $\xi, -1/\kappa$ & $\mathbb{I}(\xi =0)$ \\
GPD & $f_{GP}(x)=f_{eGP}(x)$ with $\kappa=1$ & $ \xi \geq 0$ & $\xi, -1$ & $\mathbb{I}(\xi=0)$\\
\vspace{2mm}
Gam-Gam$^{*}$  & $ f_{GG}(x)=\begin{cases} {{\left({{c_1}/{c_2}}\right)^{{{c_1}/{2}}}}\over{B(c_1/2,c_2/2)}}  {{(x/\sigma)^{-1+{c_1}/2}}\over{\left(1+c_1 x/c_2 \sigma\right)^{{{(c_1+c_2)}/{2}}}}}, \\ \sigma= {{\mu \,B(1; \,c_1/2,c_2/2)}\over{B(c_1/c_1+2c_2 ; \,c_1/2,c_2/2)}} \end{cases}$  & $(c_1>0, c_2>0)$ & $2/c_2, -2/c_1$  & 0 \\
\end{tabular}
\end{adjustbox}
\label{tab:DiffMarksDist}
\end{table}

A broad classification of tail asymptotics is given by the value of the extreme-value tail index $\xi$.  Positive values of $\xi>0$ indicate heavy tails with slow power-law tail decay; a value $\xi=0$ implies an exponential-type tail decay but with a large variety of possible subasymptotic structures; and $\xi<0$ characterizes distributions with finite upper bound and polynomial tail decay towards this bound. More refined classifications are possible, see \citet{engelke2019extremal} for an overview.  We write $\xi_L$ and $\xi_U$ to specify the lower and upper tail index, respectively.  We consider distributions with support on $[0,\infty)$, which requires $\xi_L \leq 0$ and $\xi_U \geq 0$. If we exclude the boundary cases of  tail indices being  $0$, the generalized Pareto (GP) upper $(y\to\infty)$ and lower $(y\to 0)$ tail behavior may be formally defined by the following tail expansions, respectively:
$$
1-F(y) = \ell^U(y) y^{-1/\xi_U}, \,y\to\infty, \qquad F(y) = \ell^L(1/y) y^{-1/\xi_L}, \, y\to 0,
$$
where $\ell^U(y)$ and $\ell^L(y)$ are slowly varying at infinity, i.e., $\ell^U(ty)/\ell^U(t)\to 1$ as $t\to\infty$, and similarly for $\ell^L(y)$.
Within the particularly large class with $\xi=0$, both light and heavy tails are possible, and a relevant refinement is given by the Weibull tail behavior defined by survival functions of the form
$$1-F(y) = \ell^U(y) y^{\alpha}\exp\left(-\gamma_{Wb} y^{\kappa_U}\right), \quad \alpha\in\mathbb{R}, \quad \kappa_{U},\gamma_{Wb}>0,$$
where $\kappa_{U}$ is the upper-tail  \emph{Weibull index}, with heavy tails for $\kappa_{U}<1$ and light tails for $\kappa_{U} \geq 1$; exponential tails with $\kappa_{U}=1$ form the boundary between these two scenarios. 

Table~\ref{tab:DiffMarksDist} lists several mark distribution families with their reparametrized density, lower and upper tail indices, as well as the Weibull upper-tail index, which we set to $0$ if tails are heavier than the Weibull class. The Gamma-Gamma (Gam-Gam), extended-GPD (ext-GPD) and Burr distributions are particularly flexible subasymptotic distributions with separate parameters to control the lower tail, the bulk, and the right tail of the distribution. Another flexible example is the generalized-Gamma (gen-Gam) distribution; despite its exponential upper tail ($\xi_U=0$), the parameters $\kappa$ and $c$ can allow for efficient modeling of moderate to large observations and of differences in tail heaviness. The log-Gamma (l-Gamma) and generalized Pareto distributions (GPD) are examples of parsimonious heavy-tailed models having only two parameters with positive tail index ($\xi_U>0$), but they lack flexibility in the lower tail or the bulk. The Weibull distribution is also relatively flexible  in capturing behavior in small to large observations, but it has $\xi_U=0$ and  only two parameters, thus lacking flexibility compared to the Burr, Gam-Gam, and ext-GPD models. 
The Gamma and log-Normal (l-Normal) distributions are examples of two classical models with low flexibility in the upper tail $(\xi_U=0)$, 
though their simplicity  could make them statistically more robust for prediction. For our real data application in \S\ref{sec:Chapter5Application}, we use all these distributions as candidate mark distributions in our models, comparing their performances on our landslides dataset.
 
\subsection{An alternative joint model using areal data}
\label{subsec:ArealModel}
Using the general formulation \eqref{eq:GenModel} of marked point processes, several alternative models may be constructed at different resolution levels. 
A relevant option is to define  landslide counts, sizes and observed fixed effects at low resolution using the slope units. In model \eqref{eq:GenModel}, the spatial random effects $\bm W_1$ and $\bm W_2$ are already defined at the slope-unit level rather than at pixel level. Landslide counts and sizes aggregated to slope-unit resolution  are obtained by adding up their pixel-based values within each slope unit, and similarly, fixed effect covariates at SU level are obtained by averaging them over the corresponding pixels.  This model is obtained from the general model \eqref{eq:GenModel} through the following changes. The index $j$ is removed since it is always equal to one; indices $i$ and $l$ are treated as the same index, and we keep only index $i$ for notation, and  $n_1=n_2$. The variables  $Y(\bm s_i)$, $A(\bm s_i)$, $Z_{1p}(\bm s_i)$, and $Z_{2q}(\bm s_i)$ are all defined at slope-unit level by adding up or averaging their pixel-based values. We set $A(\bm s_i)$ as missing ($\rm{NA}$) whenever $Y(\bm s_i)=0$. Finally, $a_{1r}(\bm s_i)=1$ if $r=i$, and $a_{1r}(\bm s_i)=0$ otherwise; and $a_{2r}(\bm s_l)=a_{1r}(\bm s_i)$.

This model is no longer a marked point process due to its aggregation of the marks within each slope unit, such that  information about individual points is lost. The structure of this model also differs from the approach of simply integrating predictions of model \eqref{eq:GenModel} over slope units, since covariates have also been aggregated in the new model. It is a joint areal model (i.e., a so-called lattice or network model) for landslide counts and sizes and could still provide valuable insights into their joint behavior despite its lower resolution.

\section{Bayesian inference}
\label{sec:BayesInfSim}
Hierarchical model constructions suggest using Bayesian inference. Our proposals belong to the class of latent Gaussian models, and hence approximate Bayesian estimation, such as the integrated nested Laplace approximation \citep[INLA,][]{rue2009approximate}, could be used at relatively high spatial resolution under mild conditions on the mark distributions. However,  we use several mark distributions that are not available in off-the-shelf implementations (e.g., the \texttt{R-INLA} software), or which do not necessarily comply with the requirements of INLA. Hence, instead of INLA, we use simulation-based inference based on efficient Markov chain Monte Caro (MCMC) sampling.  

\subsection{Markov chain Monte Carlo sampler}
\label{subsec:mcmc}
Let $\bm Y=\{Y(\bm s_1),\ldots,Y(\bm s_{n_1})\}^T$ denote the random vector of counts and $\bm \eta=\{\eta(\bm s_1),\ldots,$ $\eta(\bm s_{n_1})\}^T$ the corresponding random vector of log-intensities. Let $\bm U=\{l:Y(\bm s_l)>0\}$ denote the set of indices (pixels for the model in \S\ref{subsec:MarkedPoint-pixel}, slope units for the model in \S\ref{subsec:ArealModel}) for which the mark process $A$ is defined, with $n_2$ the length of vector $\bm U$. Then, $\bm A=\{\bm A(\bm s_{U_1}),\ldots, \bm A(\bm s_{U_{n_2}})\}^T$ is the vector of landslide sizes, and  $\bm \mu=\{\bm \mu(\bm s_{U_1}),\ldots, \bm \mu(\bm s_{U_{n_2}})\}^T$ is the corresponding log-median random vector. 

We use conjugate priors whenever possible. We assume independent Gaussian priors with mean zero and variance $100$ for the covariate coefficients $\beta_{1p}$, $p=1, \ldots, P_1$, $\beta_{2q}$, $q=1,\ldots,P_2$, intercepts $\gamma_1$ and $\gamma_2$, and sharing parameter $\beta$. We set relatively informative gamma hyperpriors for the precision parameters $\kappa_{\bm \eta}$, $\kappa_{\bm \mu}$, $\kappa_{\bm w_1}$ and $\kappa_{\bm w_2}$, where the parameters of the hyperpriors are chosen such as to place equal emphasis on the  variances of the  random effects with and without spatial correlation. We assume gamma hyperpriors with shape $0.25$ and rate $3$ for $\kappa_{\bm \eta}$ and $\kappa_{\bm \mu}$. For the precision parameters of spatial random effects, we set gamma hyperpriors with shape $0.25$ and rate $3/(\bar{m}\,0.7^2)$, where $\bar{m}$ is the average number of neighbors across the whole region (i.e., the average number of adjacent slope units to a given slope unit); see \cite{bernardinelli1995bayesian} for details. For the hyperparameters in $\bm \Theta_{\bm A}$, we set independent gamma priors with shape $0.25$ and rate $0.25$. 

The hierarchical joint occurrence-size model \eqref{eq:GenModel} is now fully specified as 
\begin{align}
\label{eq:HierGenModel}
Y(\bm s_i) \mid \eta(\bm s_i) &\stackrel{\rm {ind}}{\sim}  \text{Poisson}[e_i\exp\{\eta(\bm s_i)\}],\quad i=1,\ldots,n_1; \\
 \bm \eta \mid \gamma_1,\bm \beta_1,\kappa_{\bm\eta},\bm W_1 &\sim \mathcal{N}_{n_1}(\gamma_1 \bm 1+ \bm Z_{1}\bm\beta_1 + \bm A_1\bm W_1,\, \kappa_{\bm\eta}^{-1} \bm I_{n_1}); \nonumber \\
 \bm W_1\mid \kappa_{\bm w_1} &\sim \mathcal{N}_{N-1}(\bm 0,\, \kappa_{\bm w_1}^{-1} Q^{-1}) ; \quad (\bm\beta_1^T,\gamma_1)^T \sim \mathcal{N}_{P_1+1}(\bm 0,\, 100 \bm I_{P_1+1}) ; \nonumber \\ \kappa_{\bm \eta}&\sim {\rm Gamma}(0.25, 3), \quad \kappa_{\bm w_1}\sim {\rm Gamma}\{0.25, 3/(\bar{m}\,0.7^2)\};  \nonumber \\
A_{j}(\bm s_l) \mid \mu_{j}(\bm s_l) &\stackrel{\rm {ind}}{\sim} f_A[\cdot;\,\exp\{\mu_{j}(\bm s_l)\},\bm\Theta_A],\quad l:Y(\bm s_l)>0,  j=1,\ldots,Y(\bm s_l); \nonumber \\
\bm\mu \mid\gamma_2,\bm \beta_2,\bm W_1,\bm W_2,\beta, \kappa_{\bm\mu} &\sim \mathcal{N}_{L} (\gamma_2 \bm 1 + \bm Z_2 \bm\beta_{2}+ \beta\bm A_2 \bm W_1 + \bm A_2 \bm W_2, \,\kappa_{\bm\mu}^{-1} \bm I_{L}), \quad L =\sum_{i=1}^{n_1}Y(\bm s_i) ; \nonumber \\
\bm W_2\mid \kappa_{\bm w_2} &\sim \mathcal{N}_{N-1}(\bm 0,\, \kappa_{\bm w_2}^{-1} Q^{-1}) ; \quad (\bm\beta_2^T,\gamma_2,\beta)^T \sim \mathcal{N}_{P_2+2}(\bm 0,\, 100 \bm I_{P_2+2}) ; \nonumber \\ \quad \kappa_{\bm \mu}&\sim {\rm Gamma}(0.25, 3) , \quad \kappa_{\bm w_2}\sim {\rm Gamma}\{0.25, 3/(\bar{m}\,0.7^2)\},\nonumber
\end{align}
where $\bm A_1$ and $\bm A_2$ are the projection matrices of dimensions $n_1\times N$ and $L\times N$, with the $i^{th}$ and $U_{l}^{th}$ row vectors given by $\bm a_1(\bm s_i)$ and $\bm a_2(\bm s_{U_l})$, respectively; $\mathcal{N}_{d}(\bm \mu, \bm \Sigma)$ denotes the multivariate Gaussian distribution with mean vector $\bm \mu$ and covariance matrix $\bm \Sigma$; Gamma$(a,b)$ denotes the gamma distribution with shape $a$ and rate $b$; $\bm Z_1$ and $\bm Z_2$ are the design matrices of dimension $n_1\times P_1$ and $L\times P_2$ corresponding to the $P_1$- and $P_2$-dimensional vectors of covariate coefficients $\bm \beta_1$ and $\bm \beta_2$, respectively.

We update the model parameters $\kappa_{\bm \eta}$, $\kappa_{\bm \mu}$, $\kappa_{\bm w_1}$, $\kappa_{\bm w_2}$, $\bm \beta_1$, $\bm \beta_2$, $\gamma_1$, $\gamma_2$, $\beta$, $\bm W_1$, and  $\bm W_2$ using the standard Gibbs sampling algorithm \citep{casella1992explaining} by exploiting the analytically available conditional distributions of these parameters.  For hyperparameters in $\bm \Theta_{\bm A}$, we implement a standard Metropolis algorithm \citep{metropolis1953equation}.  The full conditional distributions of $\bm \eta$ and $\bm \mu$ do not have closed-form expressions, and hence we use the adaptive Metropolis adjusted Langevin algorithm (MALA) proposed in \cite{yadav2019spatial} to efficiently simulate from the full conditionals of both $\bm \eta$ and $\bm \mu$ in two separate blocks. 

Let $\bm W_1=(W_{11}, \ldots,W_{1N})^T$ and $\bm W_2=(W_{21}, \ldots,W_{2N})^T$ be the  random vectors of latent spatial effects having ICAR prior. We center their joint distributions  at zero and write $\kappa_{\bm w_1}$ and $\kappa_{\bm w_2}$, respectively, for marginal precisions. We rewrite their densities using  the pairwise difference representation:
\begin{align}
\label{eq:ICAR_Diff}
\pi(\bm W_h) \propto \exp\left\{ {{-\kappa_{\bm w_h}}\over{2}} \sum_{i \sim j}(W_{hi}-W_{hj})^2 \right\},\qquad h=1,2.
\end{align}
The joint density of the ICAR random vectors given by \eqref{eq:ICAR_Diff} leads to very moderate computational burden.  However, the pairwise difference is non-identifiable, i.e., any constant added to $\bm W_h$ cancels out in terms $W_{hi}-W_{hj}$, hence the intercept terms $\gamma_1$ and $\gamma_2$ are non-estimable. To avoid the issue of non-identifiability with non-estimable intercepts, we add the constraint  $\sum_{i=1}^{N}W_{hi}=0, h=1,2$, to center the latent spatial effects. In our MCMC algorithm, this is achieved by centering the Markov chains of both $\bm W_1$ and $\bm W_2$ at every iteration by subtracting their means, which is called hard-centering. Another way of centering the ICAR component is by assigning a Gaussian prior with mean 0 and small variance (e.g., 0.01) to the average $\sum_{i=1}^{N}W_{hi}/N, h=1,2$, hence instead of summing exactly to zero, this method ``soft-centers'' the mean by keeping it close to zero \citep{morris2019bayesian}. Yet another way of imposing the constraint is used by the INLA implementation with nondegeneracy achieved by conditioning the Gaussian vector to sum to zero.

\subsection{Simulation study and sanity check}
\label{subsec:SimStudy}
 We conducted several simulation experiments to check the estimation and prediction performance of our MCMC sampler detailed in \S\ref{subsec:mcmc}. For conciseness, we report only the results for a simulation scenario  similar to the data application in \S\ref{sec:Chapter5Application}, where we use a log-Normal size distribution for which we can further compare results across our MCMC sampler and the INLA method. We simulate data from the marked point process model \eqref{eq:GenModel} for a fixed parameter configuration given below. We use the adjacency structure of the slope units in the study area presented in \S \ref{sec:DataDescr}  to simulate the spatially correlated random effects $\bm W_1$ and $\bm W_2$ with
$\kappa_{\bm w_1}=2$ and $\kappa_{\bm w_2}=2$, and we set the precisions of spatially independent effect to $\kappa_{\bm \eta}=\kappa_{\bm \mu}=3$. Fixed-effect design matrices $\bm Z_1$ and $\bm Z_2$ are filled with the standardized covariates available in the landslide dataset; see \S\ref{subsec:CovDescr} (slope, TWI, VRM, profCurv, planCurv, TPI, LR, s--height, v--depth), and the corresponding covariate coefficients are set to $\bm \beta_1=(0.2, 0.2, 0.2, -0.2, -0.2, -0.2,  0.15,  0.15,  0.15)^T$ and $\bm \beta_2=(0.15, 0.15,  0.15, -0.1,$ $ -0.1, -0.1,  0.2,  0.2, 0.2)^T$. We use the log-Normal distribution as the mark distribution for the size with precision (i.e., inverse variance of the Gaussian) set to $\kappa=5$. The sharing parameter is set to $\beta=1$. The simulation setup is exactly the same as in the data application in \S\ref{sec:Chapter5Application}, and we have $n_1=12271$ pixels,  $N=355$ slope units, $P_1=P_2=9$ covariates, and  $L=933$ simulated landslides. 
\begin{table}[t!]
\centering
\footnotesize
\renewcommand{\arraystretch}{1}
\caption{Results of the simulation study. Posterior median, posterior standard deviation (sd), and absolute bias (bias) of the fitted model \eqref{eq:GenModel}, using INLA (inla) or the MCMC sampler (mcmc).}
\vspace{1mm}
\begin{adjustbox}{max width=\textwidth}
\begin{tabular}{c|c|c|c|c|c|c|c}
 Parameter & True value & Median (mcmc) & Median (inla) & sd (mcmc) & sd (inla) & Bias (mcmc) & Bias (inla)\\
\hline
$\kappa$ & 5 & 5.188 & 9.207 & 3.629 & 3.025 & 0.188 & 4.207\\
$\kappa_{\bm w_1}$ & 2 & 2.164 & 2.200 & 0.248 & 0.254 & 0.164 & 0.200\\
$\kappa_{\bm w_2}$ & 2 & 1.695 & 1.776 & 0.345 & 0.369 & 0.305 & 0.224\\
$\kappa_{\bm \eta}$ & 3 & 2.959 & 3.309 & 0.140 & 0.161 & 0.041 & 0.309\\
$\kappa_{\bm \mu}$ & 3 & 3.111 & 2.612 & 1.178 & 0.177 & 0.111 & 0.388\\
$\gamma_1$ & -0.5 & -0.522 & -0.507 & 0.015 & 0.015 & 0.022 & 0.007\\
$\gamma_2$ & 0.5 & 0.540 & 0.537 & 0.036 & 0.035 & 0.040 & 0.037\\
$\beta$ & 1 & 1.148 & 1.179 & 0.129 & 0.119 & 0.148 & 0.179\\
$\beta_{1_{\rm slope}}$ & 0.2 & 0.198 & 0.198 & 0.013 & 0.013 & 0.002 & 0.002\\
$\beta_{1_{\rm TWI}}$ & 0.2 & 0.177 & 0.178 & 0.011 & 0.011 & 0.023 & 0.022\\
$\beta_{1_{\rm VRM}}$ & 0.2 & 0.209 & 0.206 & 0.011 & 0.010 & 0.009 & 0.006\\
$\beta_{1_{\rm profCurve}}$ & -0.2 & -0.206 & -0.197 & 0.011 & 0.009 & 0.006 & 0.003\\
$\beta_{1_{\rm planCurve}}$ & -0.2 & -0.199 & -0.203 & 0.010 & 0.011 & 0.001 & 0.003\\
$\beta_{1_{\rm TPI}}$ & -0.2 & -0.214 & -0.212 & 0.010 & 0.01 & 0.014 & 0.012\\
$\beta_{1_{\rm LR}}$ & 0.15 & 0.129 & 0.125 & 0.019 & 0.021 & 0.021 & 0.025\\
$\beta_{1_{\rm s-height}}$ & 0.15 & 0.158 & 0.159 & 0.010 & 0.011 & 0.008 & 0.009\\
$\beta_{1_{\rm v-depth}}$ & 0.15 & 0.163 & 0.164 & 0.011 & 0.011 & 0.013 & 0.014\\
$\beta_{2_{\rm slope}}$ & 0.15 & 0.162 & 0.161 & 0.028 & 0.027 & 0.012 & 0.011\\
$\beta_{1_{\rm TWI}}$ & 0.15 & 0.153 & 0.153 & 0.027 & 0.026 & 0.003 & 0.003\\
$\beta_{2_{\rm VRM}}$ & 0.15 & 0.144 & 0.145 & 0.028 & 0.027 & 0.006 & 0.005\\
$\beta_{2_{\rm profCurve}}$ & -0.1 & -0.137 & -0.142 & 0.027 & 0.027 & 0.037 & 0.042\\
$\beta_{2_{\rm planCurve}}$ & -0.1 & -0.141 & -0.137 & 0.027 & 0.026 & 0.041 & 0.037\\
$\beta_{2_{\rm TPI}}$ & -0.1 & -0.113 & -0.115 & 0.027 & 0.025 & 0.013 & 0.015\\
$\beta_{2_{\rm LR}}$ & 0.2 & 0.162 & 0.161 & 0.045 & 0.047 & 0.038 & 0.039\\
$\beta_{2_{\rm s-height}}$ & 0.2 & 0.212 & 0.214 & 0.027 & 0.028 & 0.012 & 0.014\\
$\beta_{2_{\rm v-depth}}$ & 0.2 & 0.237 & 0.237 & 0.028 & 0.027 & 0.037 & 0.037
\end{tabular}
\end{adjustbox}
\label{tab:SimResults}
\end{table}

We use our MCMC sampler with prior distributions of hyperparameters as discussed in \S\ref{subsec:mcmc} for the inference of the model \eqref{eq:GenModel}.  We generate 250,000 MCMC samples and discard 187,500 samples as burn-in, and all summary statistics are based on the last 62,500 samples, resulting in 625 samples after thinning by a factor 100 to strongly reduce  autocorrelation in MCMC series. The Markov chains (see Figure~1 in Supplementary Material) for all hyperparameters and latent parameters visibly  converge after about 50,000 samples. 
Table~\ref{tab:SimResults} compares the estimation performance of INLA and our MCMC sampler. In general, the absolute bias and standard error in the parameter estimates are comparable in both approaches,  especially for fixed effect parameters. The uncertainties in hyperparameter estimates are comparable except for the precision parameter $\kappa_{\bm \mu}$ where the MCMC sampler is more variable.
On the other hand, INLA strongly overestimates the precision parameter ($\hat{\kappa}=9.207$) of log-normal observations, but again this seems acceptable given that the estimation uncertainty is relatively high due to the smaller sample size.  

We also assess the prediction performance of our MCMC sampler by an out-of-sample (OOS) experiment using 15-fold cross-validation, where we randomly partition the  355 slope units into 15 different sets. We repeat this step to obtain 15 separate combinations, each of which consists of 355 slope units with data at one set of slope units treated as missing, and the other 14 sets of slope unit data used for training and to predict data at the missing slope units. Figure~1 in the Supplementary Material shows the OOS prediction performance of our MCMC sampler; its first row reports the true (left display)  and predicted (right display) landslide counts, and by analogy the second row shows true and predicted landslide sizes; all  plots are in log-scale. We conclude from these plots that our MCMC sampler performs satisfactorily for predicting landslide counts and sizes jointly at unobserved slope units.

\section{Data application}
\label{sec:Chapter5Application}

%
\subsection{Results for the marked point process models}
\label{subsec:Chapter5ApplicationPoint}
We fit the proposed model \eqref{eq:GenModel} to the landslides data detailed in \S\ref{sec:DataDescr} and jointly model landslide counts and sizes for different choices of mark distribution; see Table~\ref{tab:DiffMarksDist} for the details of mark distributions. For each mark distribution, we considered eight different submodels possessing distinct properties while being nested within the general model \eqref{eq:GenModel}. These models share the same general structure but have different specifications, i.e., whether or not they include the random effects $\varepsilon_{\bm\eta}$ and $\varepsilon_{\bm\mu}$ in \eqref{eq:GenModel}, and whether or not the sharing parameter $\beta$ is fixed to zero (base model), corresponding to landslide occurrences and sizes being independent of each other;  see Table~\ref{tab:DiffModels} for details on the different submodels. 
\begin{table}[t!]
\centering
\tabcolsep=0.11cm
\renewcommand{\arraystretch}{1}
\caption{Candidate models that differ with respect to the inclusion of independently and identically distributed (iid) random effects (indicated through different subscripts) and to the sharing coefficient being fixed at $\beta=0$ in the general model \eqref{eq:GenModel} (indicated through a superscript of $0$).}
\begin{adjustbox}{max width=\textwidth}
\begin{tabular}{c|c|c|c}
 Model & iid random effect $\varepsilon_{\bm\eta}$ in $\bm \eta(\bm s)$ & iid random effect $\varepsilon_{\bm\mu}$ in $\bm \mu(\bm s)$ & $\beta=0$ (base model) \\
 \hline
  M$_1$ & yes & yes & no \\ 
 M$_2$ & no & no & no \\ 
 M$_3$ & yes & no & no \\ 
 M$_4$ & no & yes & no \\ 
  M$_1^0$ & yes & yes & yes \\ 
 M$_2^0$ & no & no & yes \\ 
 M$_3^0$ & yes & no & yes \\ 
 M$_4^0$ & no & yes & yes \\ 
\end{tabular}
\end{adjustbox}
\label{tab:DiffModels}
\end{table}
The full model presented in \eqref{eq:GenModel} is noted M$_1$; it includes iid random effects in both latent processes, whereas the base model M$_1^0$ is similar to model M$_1$ but with fixed $\beta=0$. Model M$_2$ only includes the spatially structured random effects in both latent processes but does not have any iid random effects, and model M$_2^0$ is the same as model M$_2$ but with fixed $\beta=0$. We use all  available covariates detailed in \S\ref{subsec:CovDescr}, along with the total area of slope units that is only included as an additional covariate for landslide sizes.

We fit all candidate models using the MCMC sampler detailed in \S\ref{subsec:mcmc}, generating 100,000 posterior samples for each model. All summary statistics reported below are based on the last 25,000 MCMC samples after deleting the first 75,000 burn-in samples. The Markov chains (see \S2.2 in the Supplementary Material) converge fairly quickly to a stationary distribution, and appear to mix well for all hyperparameters and latent parameters. For fitting models M$_2$, M$_3$ and M$_4$, we actually fit model M$_1$ but with very high fixed precision parameters $\kappa_{\bm\eta}=1000$ and $\kappa_{\bm \mu}=1000$ whenever the corresponding independent random effect is absent, and their base model counterparts are fitted with additionally fixing the sharing parameter $\beta$ to $0$. 
The choice of hyperparameter priors is detailed in \S\ref{subsec:mcmc}.

For comparing the goodness-of-fit and predictive performance of candidate models, we design three types of experiments, providing within-sample (WS) or out-of-sample (OOS) diagnostics. OOS analyses are based either on a 10-fold cross-validation where we remove data to be predicted for entire slope units, or on a 5-fold thinning-based approach  \citep{leininger2017bayesian} where we remove individual landslides from the dataset, such that the training and validation data can contain landslides from the same slope unit. 
In the 5-fold thinning-based OOS approach, we randomly remove $20\%$ of landslides and their associated sizes in each fold, and predict them in our MCMC algorithm. We determine the fold allocation of each landslide by sampling from a uniform random variable for each landslide and then categorize the landslides as part of fold $i\in\{1,2,...5\}$  if the uniform value is in $[(i-1)\times 0.2, i \times 0.2]$.  This randomly divides the total number of 933 landslides into five folds of sizes 188,183,172,199,191. After estimating the point process intensity on the remaing 4/5 of data, we re-adjust the estimated intensity by multiplying it by 5/4.
All  models are fitted at the pixel level with random effects defined at slope-unit level, but we report prediction results at the slope unit level, which is easier to interpret and compare. Results at the slope-unit level are obtained by aggregating the predicted counts and sizes from the pixel level.
Subsequently, we use the component-wise mean of $\exp(\bm\mu_{\rm post})$ and $\exp(\bm\eta_{\rm post})$ as  posterior predictive estimates for the landslide counts and sizes, respectively, where $\bm\eta_{\rm post}$ and $\bm \mu_{\rm post}$ are the posteriors samples of latent parameters corresponding to counts and sizes, respectively. In WS and 10-fold OOS experiments, we directly impute the sample of $\bm\eta_{\rm post}$ and $\bm \mu_{\rm post}$ at the prediction locations, whereas for the thinning-based OOS approach we use the posterior samples of $\gamma_1, \gamma_2, \bm\beta_1, \bm\beta_2, \bm W_1$ and $\bm W_2$ to recreate the posterior samples of $\bm\eta_{\rm post}$ and $\bm \mu_{\rm post}$ using the model construction \eqref{eq:HierGenModel}. For conciseness, we report only results for 10-fold OOS experiments, whereas the results for the WS and 5-fold thinning-based approach are detailed in \S2.3 of the Supplementary Material.

\begin{table}[t!]
\centering
\renewcommand{\arraystretch}{1}
\caption{Comparison of OOS predictions in the data application reporting Area-Under-the-Curve (AUC) and mean absolute error (AbsError) diagnostics for models M$_1$, M$_1^0$, M$_2$ and M$_2^0$ with different mark distributions listed in Table~\ref{tab:DiffMarksDist}, obtained for 10-fold cross-validation. Bold blue-colored digits highlight the best performance among different models and different marks, and light-blue digits show the best performance for a particular choice of mark distribution.  Values reported as NA correspond to non-applicable AUC due to numerical instabilities arising with predicted values.}
\vspace{1mm}
\begin{adjustbox}{max width=\textwidth}
\begin{tabular}{c|c|c|c|c|c|c|c|c|c|c|c}
Summary & Model & Data & $f_{Bu}$ & $f_{G}$ & $f_W$ & $f_{lG}$  & $f_{lN}$ & $f_{eGP}$ & $f_{GP}$ &$f_{GG}$& $f_{gG}$\\
\hline
\multirow{8}{*}{AUC} & \multirow{2}{*}{M$_1$} & count & 0.771 &  \cl{blue}{0.814} &  \cl{blue}{0.797} &  \bcl{blue}{0.818} &  \cl{blue}{0.797} &  \cl{blue}{0.802} & 0.78 &  \cl{blue}{0.786} &  \cl{blue}{0.791}\\


 &  & size & 0.792 &  \bcl{blue}{0.837} & 0.760 & 0.632 &  \cl{blue}{0.813} & 0.777 & 0.79 & 0.802 &  \cl{blue}{0.825}\\

\cline{2-12}

 & \multirow{2}{*}{M$_1^0$} & count & 0.786 & 0.795 & 0.795 & 0.803 & 0.782 & 0.786 &  \cl{blue}{0.798} & 0.779 & 0.786\\


 &  & size & \cl{blue}{0.794} & 0.775 &  \cl{blue}{0.768} &  \cl{blue}{0.646} & 0.798 &  \cl{blue}{0.804} &  \cl{blue}{0.802} &  \cl{blue}{0.821} & 0.763\\

\cline{2-12}

 & \multirow{2}{*}{M$_2$} & count & 0.786 & 0.787 & 0.787 & 0.785 & 0.787 & 0.786 & 0.786 & 0.786 & 0.787\\


 &  & size & 0.617 & 0.618 & 0.626 & 0.636 & 0.715 & 0.618 & 0.627 & NA & 0.617\\

\cline{2-12}

 & \multirow{2}{*}{M$_2^0$} & count &  \cl{blue}{0.787} & 0.786 & 0.786 & 0.785 & 0.786 & 0.787 & 0.786 & 0.785 & 0.787\\


 &  & size & 0.622 & 0.622 & 0.642 & 0.645 & 0.809 & 0.626 & 0.651 & NA & 0.62\\

\hline

\multirow{8}{*}{AbsError} & \multirow{2}{*}{M$_1$} & count & 1.59 &  \cl{blue}{1.56} &  \cl{blue}{1.56} & 1.91 &  \cl{blue}{1.56} &  \cl{blue}{1.56} & 1.60 &  \bcl{blue}{1.55} &  \cl{blue}{1.58}\\


 &  & size &  \cl{blue}{130} & 142 & 147 & 277 & 133 & \bcl{blue}{128} &  \cl{blue}{131} & 132 & 149\\

\cline{2-12}

 & \multirow{2}{*}{M$_1^0$} & count &  \cl{blue}{1.58} & 1.56 & 1.56 &  \cl{blue}{1.58}& 1.57 & 1.58 &  \cl{blue}{1.58} & 1.55 & 1.58\\


 &  & size & 135 &  \cl{blue}{132} &  \cl{blue}{146} &  \cl{blue}{264} &  \cl{blue}{131} & 133 & 134 &  \cl{blue}{132} &  \cl{blue}{145}\\

\cline{2-12}

 & \multirow{2}{*}{M$_2$} & count & 16.0 & 16.2 & 16.2 & 16.2 & 15.7 & 16.0 & 16.2 & 15.7 & 15.8\\


 &  & size & 297 & 292 & 281 & 277 & 178 & 293 & 283 & 303 & 297\\

\cline{2-12}

 & \multirow{2}{*}{M$_2^0$} & count & 15.8 & 15.4 & 15.4 & 16.3 & 16.2 & 15.8 & 15.4 & 15.7 & 15.8\\


 &  & size & 289 & 283 & 270 & 268 & 135 & 283 & 259 & 301 & 293\\

\end{tabular}
\end{adjustbox}
\label{tab:OSD.M1}
\end{table}
Model M$_1$ outperforms models M$_2$, M$_3$, M$_4$ and its baseline counterpart; see, Tables 1, 2 and 3 in the Supplementary Material for a detailed comparison across all fitted models, and Table~\ref{tab:OSD.M1} for a summary. Table~\ref{tab:OSD.M1} reports the Area-Under-the-Curve (AUC) values and mean absolute errors in the 10-fold cross-validation experiment for models M$_1$, M$_1^0$, M$_2$ and M$_2^0$ with different mark distributions.  The AUC metric is a performance measure for binary data, and so here we first  threshold observed and predicted landslide counts and sizes to obtain binary information. The thresholds for AUC calculation are set to one for landslide counts, which assesses presence/absence of landslides, while for landslide sizes, it is set to $77.6$, the average landslide size (i.e., square root of landslide area). Here the AUC metric puts focus on how well models discriminate between small and moderately large values, whereas absolute errors are more strongly influenced by extreme values due to the heavy tails in observations of counts and sizes. 
In Table~\ref{tab:OSD.M1}, model M$_1$ with shared spatial effect generally outperforms its corresponding base model M$_1^0$ when predicting landslide counts, and it has smaller absolute errors when predicting landslide sizes using flexible sub-asymptotic mark distributions allowing for positive upper tail index $\xi_U$. In contrast, model M$_1^0$ has often smaller absolute errors for landslide sizes when other mark distributions with lighter tails are used, e.g., Gamma, Weibull, log-Normal, and generalized-Gamma; however, these mark distributions have higher absolute errors compared to M$_1$ for predicting landslide counts. When comparing the different models (M$_1$, M$_1^0$, M$_2$, and M$_2^0$) and different mark distributions listed in Table~\ref{tab:DiffMarksDist}, the overall best performing model is M$_1$, but there is no clear winner for the mark distribution. With model M$_1$, the Gamma distribution has the highest AUC value (0.837) for landslide sizes; the log-Gamma distribution has the highest AUC value (0.818) for landslide counts; the  extended-GPD distribution has the lowest absolute error (128) for landslide sizes; and the Gamma-Gamma distribution has the lowest absolute error (1.55) for landslide counts.
The three sub-asymptotic distributions (ext-GPD, Burr, and Gamma-Gamma) have the first, second, and fourth smallest absolute errors, respectively, for predicting landslide sizes, often much smaller as for other mark distributions; the highest (i.e., worst) absolute error is obtained with the log-Gamma (277) distribution. We also compared the different models (M$_1$, M$_1^0$, M$_2$, M$_2^0$) and different mark distributions using the continuous ranked probability score \citep[CRPS;][]{gneiting2007strictly} and the deviance information criterion \citep[DIC;][]{gelman2014understanding}, leading to very similar interpretations; see Table~4 in the Supplementary Material. Overall, Model M$_1$ with Gamma-Gamma marks consistently provides a very good performance in predicting landslide counts and sizes.

In the WS setting (see Tables 1, 2 and 3 in the Supplementary Material), the simpler models with Gamma, Weibull and log-Normal marks possessing relatively light upper tails outperform the more flexible and complex models; however in the OOS setting, we see the opposite behavior, and flexible sub-asymptotic mark distributions, such as Gamma-Gamma, ext-GPD and Burr models, outperform the more parsimonious classical models. Based on our cross-validation experiments, we select model M$_1$ with Gamma-Gamma mark distribution as the best model overall for jointly predicting landslide counts and sizes at unobserved slope units. A similar interpretation follows from thinning-based cross-validation; see Table~5 in the Supplementary Material.

\begin{figure}[t!]
\includegraphics[width=1\textwidth]{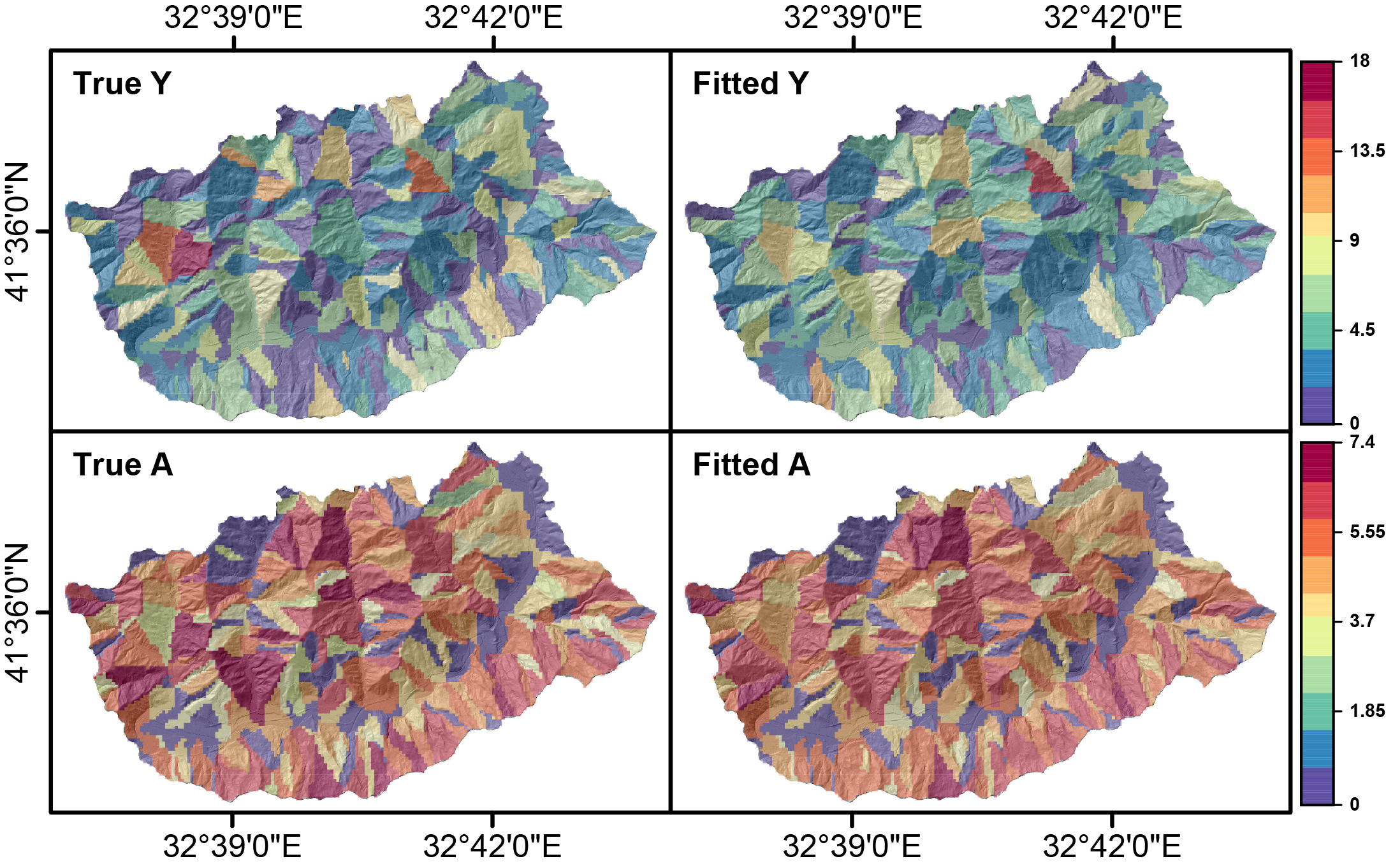}
\caption{Fitted vs.\ true landslide counts and sizes for the landslides data detailed in \S\ref{sec:DataDescr}, where the fitted data are obtained based on model M$_1$ with Gamma-Gamma marks using 10-fold cross-validation. First row: true (left)  and predicted landslide counts (right). Second row: true (left) and predicted landslides sizes (right), both on log-scale.}
\label{fig:WSD.W.OSD.GG}
\end{figure}
Figure~\ref{fig:WSD.W.OSD.GG} shows the posterior predictive means (second column) of landslide counts (top row) and landslide sizes (bottom row) when assuming a Gamma-Gamma mark distribution in the 10-fold cross-validation setting. Results look visually satisfactory, and we conclude from these plots that our proposed model accurately predicts both observed and unobserved landslide counts and sizes at the slope unit level.

\begin{table}
\centering
\renewcommand{\arraystretch}{1}
\caption{ Posterior median, standard error (se) and $95\%$  credible interval lower bound (CI$^{-}$) and upper bound (CI$^{+}$) for model M$_1$ with Gamma-Gamma mark distribution in the WS setting.}
\vspace{1mm}
\begin{adjustbox}{max width=\textwidth}
\begin{tabular}{c|c|c|c|c|c|c|c|c|c}
 parameter & median & se & CI$^{-}$ & CI$^{+}$ &   parameter & median & se & CI$^{-}$ & CI$^{+}$\\
\hline

$c_1$& 32.508 & 7.580 & 20.441 & 48.292 & $\beta_{1_{\rm TPI}}$& 0.029 & 0.035 & -0.037 & 0.099\\


$c_2$ & 8.254 & 1.346 & 6.630 & 11.731 & $\beta_{1_{\rm LR}}$ & 0.255 & 0.060 & 0.125 & 0.356\\


$\kappa_{\bm w_1}$ & 0.783 & 0.118 & 0.593 & 1.020 & $\beta_{1_{\rm s-height}}$ & -0.035 & 0.042 & -0.112 & 0.046\\


$\kappa_{\bm w_2}$ & 3.519 & 1.017 & 2.178 & 6.094 & $\beta_{1_{\rm v-depth}}$ & 0.175 & 0.029 & 0.119 & 0.234\\


$\kappa_{\bm \eta}$& 2.264 & 0.262 & 1.947 & 3.028 & $\beta_{2_{\rm slope}}$ & -0.041 & 0.024 & -0.086 & 0.012\\


$\kappa_{\bm \mu}$ & 5.630 & 1.108 & 3.857 & 8.348 & $\beta_{2_{\rm TWI}}$ & -0.018 & 0.026 & -0.064 & 0.036\\


$\gamma_1$& -3.239 & 0.035 & -3.295 & -3.156 & $\beta_{2_{\rm VRM}}$ & 0.001 & 0.026 & -0.046 & 0.048\\


$\gamma_2$ & 4.080 & 0.038 & 4.008 & 4.158 & $\beta_{2_{\rm profCurve}}$ & 0.012 & 0.025 & -0.037 & 0.058\\


$\beta$ & -0.444 & 0.093 & -0.623 & -0.261 & $\beta_{2_{\rm planCurve}}$ & -0.041 & 0.028 & -0.101 & 0.005\\


$\beta_{1_{\rm slope}}$ & 0.380 & 0.037 & 0.307 & 0.461 & $\beta_{2_{\rm TPI}}$ & 0.036 & 0.026 & -0.015 & 0.088\\


$\beta_{1_{\rm TWI}}$& 0.095 & 0.038 & 0.029 & 0.162 & $\beta_{2_{\rm LR}}$ & 0.026 & 0.042 & -0.046 & 0.111\\


$\beta_{1_{\rm VRM}}$& -0.125 & 0.037 & -0.206 & -0.067 & $\beta_{2_{\rm s-height}}$ & 0.009 & 0.029 & -0.052 & 0.061\\


$\beta_{1_{\rm profCurve}}$ & 0.058 & 0.028 & -0.006 & 0.109 & $\beta_{2_{\rm v-depth}}$ & 0.038 & 0.028 & -0.013 & 0.09\\


$\beta_{1_{\rm planCurve}}$& 0.006 & 0.043 & -0.084 & 0.074 & $\beta_{2_{\rm area-SU}}$ & 0.010 & 0.036 & -0.059 & 0.082\\

\end{tabular}
\end{adjustbox}
\label{tab:Gamma-Gamma.WSD.summary}
\end{table}

Table~\ref{tab:Gamma-Gamma.WSD.summary} shows posterior summaries for model M$_1$ with the Gamma-Gamma mark distribution in the WS setting. The estimated precision parameters for the ICAR components are $\hat{\kappa}_{\bm w_1}=0.783$ and $\hat{\kappa}_{\bm w_2}=3.519$. Moreover,  estimated precisions for the independent random effects are $\hat{\kappa}_{\bm \eta}=2.264$ and $\hat{\kappa}_{\bm \mu}=5.630$, which highlights non-negligible unstructured spatial variation in both landslide counts and sizes. The parameter $\beta$, which controls the sharing of information between models for counts and sizes, is estimated  highly negative, $\hat{\beta}=-0.444$, which shows that a small number of landslides in a particular region is positively correlated with a bigger landslide size (square root of landslide area) in the same region. The estimated upper tail index is  $(2/\hat{c}_2)=0.243$, which confirms that  landslide sizes are quite heavy-tailed. Since the model is here fitted to the square root of landslide areas, the tail index for landslide areas themselves would be about twice as large. As expected, the regression coefficient for the slope steepness, $\hat{\beta}_{1_{\rm slope}}=0.38$,  is highly positive for landslide counts and significant (with lower bound of the credible interval far from zero); however, for landslide sizes, it turns out to be non-significant. The regression coefficient for VRM is negative $\hat{\beta}_{1_{\rm VRM}}= -0.125$,  as VRM is related to the solidness of the soil. Most  covariates related to the size process seem to be non-significant; this may be due to the rather small number of samples for the landslide size, resulting in quite large uncertainties in parameter estimates. 

\begin{figure}[t!]
\centering
  \includegraphics[width=1\textwidth]{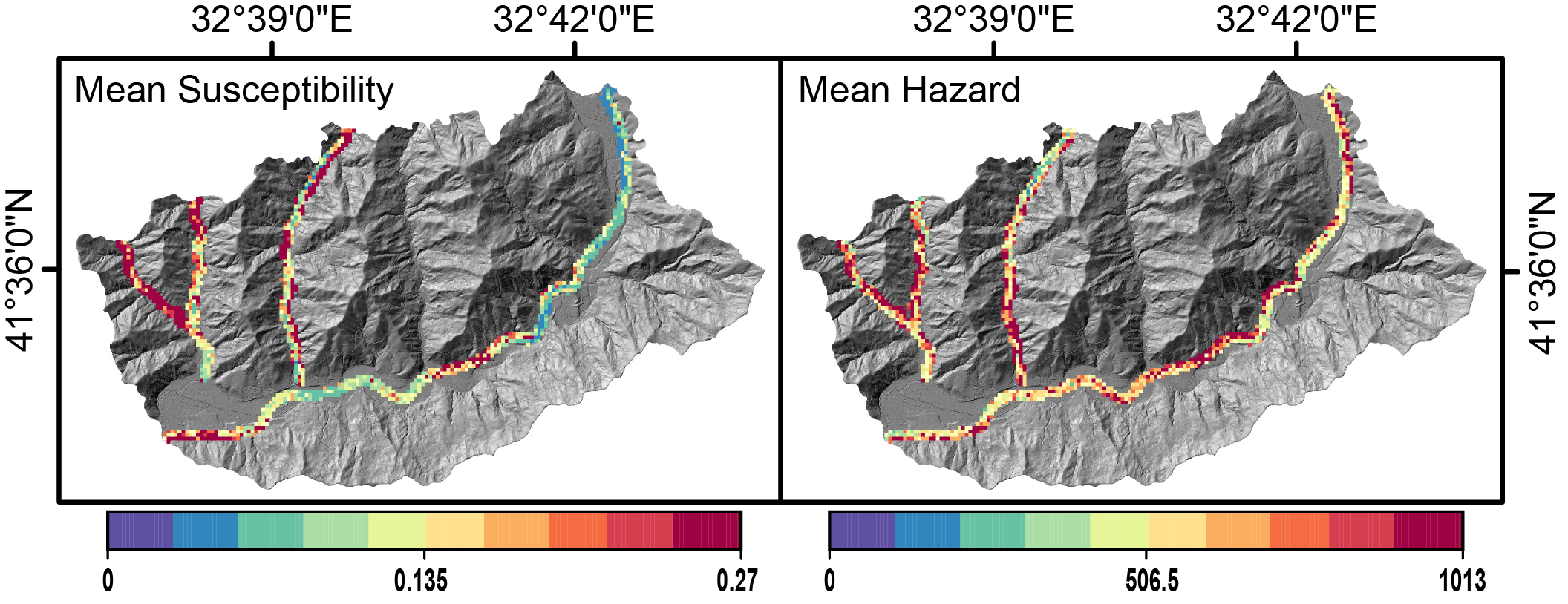}
\caption{Estimated susceptibility (left) and hazard (right) across the four main roads (colored) in the study regions based on the model M$_1$ with Gamma-Gamma marks.}
\label{fig:HazardSucep}
\end{figure}

Finally, to illustrate the practical usefulness of our joint modeling framework for landslide hazard assessment and country planning, Figure~\ref{fig:HazardSucep} shows the mean susceptibility and mean hazard plots across four different main roads in the study region, based on the best model M$_1$ with Gamma-Gamma marks. These main roads are created with a buffer of 50 meters on each side.
This allows assessing the risk associated with  landslide events close to roads, which pose a threat of human and economic losses. The susceptibility is here defined as the posterior probability of having at least one landslide, i.e., it is of the form $1-\exp\{-\exp(\bm\eta)\}$ using the Poisson assumption. By contrast, the calculation of landslide hazard includes information from both landslide counts and sizes, and we here define it as the product $\exp(\bm\eta) \times \exp(2\bm\mu)$, where $\bm\eta$ and $\bm\mu$ denote the log-intensity and log-median processes associated with landslide counts and sizes, respectively. In other words, the hazard measures the propensity of a given spatial location (conditional on predictors) to experience landslides that are both likely and big. Note that the factor 2 is used to transform the landslide sizes (square root of landslide area) back to the original landslide area scale. Estimates of susceptibility and hazard are obtained by their posterior means, averaging the respective quantities defined above after replacing $\bm\eta$ and $\bm\mu$ with their post-burn-in posterior values from the MCMC output. Note that the correlation between the count and size processes assumed in our marked point process models is key to accurately computing the landslide hazard, as the latter is a function of both. The mean susceptibility plot in the left panel in Figure~\ref{fig:HazardSucep} across four different roads shows that there is a non-zero probability of landslide occurrence close to roads. The two smallest roads appear as the most dangerous ones, and the longest roads have the lowest risk of landslide occurrence, certainly because most parts of the longest roads are in the flatter parts of the study region. 
In the mean hazard plots in the right panel of Figure~\ref{fig:HazardSucep}, many similar patterns arise, but sometimes we also discern a different behavior with opposite trends of susceptibilities and hazards (e.g., for the longest road) because the estimated sharing parameter of latent spatial fields $\hat{\beta}=-0.444$ is negative. Our joint modeling framework is beneficial in providing estimates of susceptibility and a unified way of calculating hazard-related quantities that depend on both landslide counts and sizes. 
We can also compute the aggregated hazard for the whole road network by adding up the pixel hazard over all four roads. The resulting simulation-based posterior predictive quantiles at  0$\%$, 5$\%$, 25$\%$, 50$\%$, 75$\%$, 95$\%$ and 100$\%$  are $58075$, $66114$, $78122$,  $86584$,  $95462$, $111103$ and $135005$, respectively, which shows that we can estimate this quantity with moderate uncertainty. Comparing such aggregated hazard quantities for different sub-domains of the study region could also be helpful for risk mitigation and country planning, as one might only be able to stabilize the most risky slopes with a fixed budget.

\subsection{Results for areal data: joint model at slope unit level}\label{subsec:arealresults}
 For comparison, we also fit the joint models to  aggregated areal data, as described in \S\ref{subsec:ArealModel}. 
 To make inference, we use straightforward modifications of our MCMC sampler from \S\ref{subsec:mcmc}. We generate 100,000 posterior samples, and all posterior inferences are based on the last 25,000 samples after deleting the first 75,000 burn-in samples. 

Table~6 in the Supplementary Material shows the AUC values and absolute errors for different mark distributions with models M$_1$, M$_1^0$, M$_2$, and M$_2^0$ in the WS experiment. Model M$_1$ has higher AUC  and lower absolute errors overall for both landslide counts and  sizes. Therefore, we select model M$_1$ with Burr mark distribution as the best model due to its overall better performance for both  processes.  The Weibull and log-Normal mark distributions also have very good and quite similar performance, especially in terms of  lowest absolute error for landslide sizes. Nevertheless, by analogy with  previous results from \S\ref{subsec:Chapter5ApplicationPoint}, we expect that  Weibull and log-Normal distributions will perform worse when predicting unobserved landslide data in the OOS setting.  

Table~7 in the Supplementary Material shows summary statistics for model M$_1$ with Burr mark distribution. Its  tail index for landslide sizes is estimated at $(1/\hat{c}\,\hat{\kappa})=0.087$, indicating heavy-tailed landslide sizes. The estimated precision of $\hat{\kappa}_{\bm w_1}=0.111$ for the shared ICAR random effect $\bm W_1$ is rather small, which implies that the shared spatial random effect jointly explains a large part of the variability in both processes, with a highly positive   sharing parameter $\hat{\beta}=1.443$. At the slope unit levels, a larger number of landslides therefore contributes to larger landslide sizes. Recall that this is the opposite sign of the sharing coefficient for the models fitted at the pixel level resolution in \S\ref{subsec:Chapter5ApplicationPoint}: at very small scales, occurrence of several landslides typically corresponds to relatively small landslides, whereas at the coarser slope unit level a strong multiplicity of landslides tends to go hand in hand with larger total sizes. Several of the covariate coefficients are highly significant, e.g., slope for counts, planCurve for counts and sizes, s-height for counts and sizes, and LR for counts and sizes. The variability of estimated covariate coefficients is often larger than for the pixel-level model, such that several covariates remain non-significant. This appears to be a disadvantage of modeling aggregated data directly at the slope unit level compared to the marked point process models in \S\ref{subsec:MarkedPoint-pixel}, for which data and covariate information is fully exploited at the pixel level for individual landslides rather than at the slope unit level. 

\section{Summary and conclusions}
\label{sec:Conclsuion}

We proposed a joint spatial modeling framework allowing for stochastic interactions between landslide counts and sizes. The  size is treated as a ``mark'' in a marked point process framework, and various types of sub-asymptotic mark distributions justified by extreme-value theory are used. The high flexibility of our new models suggests that they can be more generally used to jointly model count and size processes in other contexts, e.g., for wildfire counts and burnt areas, where spatio-temporal extensions could be developed. Our model is based on a hierarchical construction with  Gaussian ICAR  priors to capture spatial dependence at the latent level. The spatially structured interaction between the landslide count and size processes is measured by a shared ICAR prior with sharing coefficient  estimated from the data. The ICAR prior is used efficiently at the slope unit level, and additionally, the sparse structure of ICAR priors reduces the computation burden significantly, thus opening the door to fit our models in very high spatial dimensions.

We fitted our models using a customized MCMC, in which the high-dimensional latent variables $\bm \eta$ and $\bm \mu$ are simulated efficiently using the Metropolis adjusted Langevin algorithm (MALA). The great benefit of our MCMC implementation is that various types of mark distributions and model structures can be easily incorporated and tested, unlike other techniques such as INLA. 

To study a landslide inventory from Turkey, we compared and fitted models with structural differences in the bivariate latent process and the mark distributions. Models with shared random effects and sub-asymptotic mark distributions performed especially well for predicting unobserved data at the slope unit scale, compared to baseline models (with independent count and size processes) or models using classical mark distributions. In future research, it would be interesting to use our models in even higher spatial dimensions and to apply them to other point patterns available for very large observation windows. 
Another research avenue is towards spatio-temporal extensions for landslide counts and sizes or other environmental and ecological marked point patterns, and specifically to dynamic models with self-inflating or self-deflating effects of occurrences on the point-process intensity.

\section*{Acknowledgments}
This publication is based upon work supported by the King Abdullah University of Science and Technology (KAUST) Office of Sponsored Research (OSR) under Award No. OSR-CRG2020-4338.

\baselineskip 14pt
\bibliographystyle{CUP}
\bibliography{ref}

\baselineskip 10pt

\end{document}